\documentclass[aps,prd,reprint,twocolumn,superscriptaddress,showpacs]{revtex4-2}
\usepackage{amsfonts}
\usepackage{mathrsfs}
\usepackage{amsmath}
\usepackage{color}
\usepackage{natbib}
\usepackage{graphicx}
\usepackage{bm}
\usepackage{amssymb}
\usepackage{xspace}
\usepackage{epstopdf}
\usepackage{dcolumn}
\usepackage{longtable}
\usepackage{multirow}
\usepackage[colorlinks=true, letterpaper=true, pdfstartview=FitV, linkcolor=blue, citecolor=blue, urlcolor=blue]{hyperref}
\usepackage{float}

\makeatletter

\newcommand{\Rmnum}[1]{\expandafter\@slowromancap\romannumeral #1@}
\makeatother

\begin{document}

\title{Monolayer and bilayer PtCl$_3$: Energetics, magnetism, and band topology}

\author{Yalong Jiao}
\email{yalong.jiao@hebtu.edu.cn}
\affiliation{College of Physics, Hebei Key Laboratory of Photophysics Research and Application, Hebei Normal University, Shijiazhuang 050024, China}
\affiliation{Research Laboratory for Quantum Materials, Singapore University of Technology and Design, Singapore 487372, Singapore}

\author{Xu-Tao Zeng}
\affiliation{School of Physics, Beihang University, Beijing 100191, China}

\author{Cong Chen}
\affiliation{Department of Physics, The University of Hong Kong, Hong Kong, China}

\author{Zhen Gao}
\affiliation{College of Physics, Hebei Key Laboratory of Photophysics Research and Application, Hebei Normal University, Shijiazhuang 050024, China}

\author{Xian-Lei Sheng}
\email{xlsheng@buaa.edu.cn}
\thanks{Y. Jiao and X.-T. Zeng contributed equally to this work.}
\affiliation{School of Physics, Beihang University, Beijing 100191, China}

\author{Shengyuan A. Yang}
\affiliation{Research Laboratory for Quantum Materials, Singapore University of Technology and Design, Singapore 487372, Singapore}

\begin{abstract}
Two-dimensional (2D) magnetic materials hosting nontrivial topological states are interesting for fundamental research as well as practical applications. Recently, the topological state of 2D Weyl half-semimetal (WHS) was proposed, which hosts fully spin polarized Weyl points robust against spin-orbit coupling in a 2D ferromagnetic system, and single-layer PtCl$_3$ was predicted as a platform for realizing this state. Here, we perform an extensive search of 2D
PtCl$_3$ structures, by using the particle swarm optimization technique and density functional theory calculation. We show that the desired PtCl$_3$ phase  corresponds to the most stable one at its stoichiometry. The 2D structure also possesses good thermal stability up to 600 K. We suggest SnS$_2$ as a substrate for the growth of 2D PtCl$_3$, which has excellent lattice matching and preserves the WHS state in PtCl$_3$. We find that uniaxial strains along the zigzag direction maintain the WHS state, whereas small strains along the armchair direction drives a topological phase transition from the WHS to a quantum anomalous Hall (QAH) insulator phase. Furthermore, we study bilayer PtCl$_3$ and show that the stacking configuration has strong impact on the magnetism and the electronic band structure. Particularly, the $AA'$ stacked bilayer PtCl$_3$  realizes an interesting topological state --- the 2D antiferromagnetic mirror Chern insulator, which has a pair of topological gapless edge bands. Our work provides guidance for the experimental realization of 2D PtCl$_3$ and will facilitate the study of 2D magnetic topological states, including WHS, QAH insulator, and magnetic mirror Chern insulator states.

\end{abstract}
\pacs{}
\maketitle

\section{Introduction}

Topological states of matter have been extensively studied in nonmagnetic crystalline materials~\cite{hasan2010colloquium,RevModPhys831057,shen2012topological,bernevig2013topological,bansil2016colloquium,armitage2018weyl}. Recently, there are two emerging trends in this field. The first is to extend the study to magnetic materials~\cite{tokura2019magnetic,xu2020high,bernevig2022progress}, motivated by the possibility to control electronic topology via magnetism. The second is to explore topological materials in lower dimensions, especially in two-dimensional (2D) materials~\cite{ren2016topological,kou2017two,wang2019two,feng2021two,elcoro2021magnetic}, because their properties can be readily tailored with well developed experimental techniques, e.g., by applying lattice strain or by controlling stacking pattern (in a multi-layer structure).

The merging point of the two trends, namely, the study of 2D magnetic materials with nontrivial topology, clearly attracts a lot of interest~\cite{liu2021magnetic}. On the fundamental level, there are topological states unique to such systems. A prominent example is the quantum anomalous Hall (QAH) insulator, characterized by nontrivial Chern numbers and featuring chiral edge modes~\cite{haldane1988model}. On the practical level, these materials are  promising platforms for constructing novel electronic and spintronic devices. As mentioned, the 2D nature makes them highly tunable; and the chiral edge modes of QAH insulators can be used as transport channels with ultra-low dissipation~\cite{hou2020progress}. Experimentally, topological 2D magnetic systems were first investigated in magnetically doped topological insulator thin films~\cite{ruifirst,qahfirst}. Recently, with the discovery of 2D materials with intrinsic magnetism, such as 2D CrI$_3$~\cite{huang2017layer}, Cr$_2$Ge$_2$Te$_6$~\cite{gong2017}, VSe$_2$~\cite{bonilla2018strong}, and etc.~\cite{du2016weak,lee2016ising,Hara2018,burch2018magnetism}, there is increasing interest in exploring topological states in such systems. So far, the most studied system is the 2D MnBi$_{2n}$Te$_{3n+1}$ system, which hosts several magnetic topological insulating states~\cite{li2019intrinsic,otrokov2019unique,otrokov2019prediction,deng2020quantum}.

In a previous work~\cite{you2019two}, we proposed a 2D magnetic topological state --- the 2D Weyl half-semimetal (WHS), whose Fermi surface are consisting of twofold degenerate Weyl points belonging to a single spin channel, such that the state is simultaneously a half metal and a Weyl semimetal. The 2D WHS can be regarded as a critical state between two QAH phases with opposite Chern numbers. A concrete material, the single-layer PtCl$_3$ with the CrI$_3$-type structure, was predicted as an ideal system to realize the 2D WHS state~\cite{you2019two}, and till now, it remains to be the only material candidate.

In Ref.~\cite{you2019two}, only basic properties of single-layer PtCl$_3$ were discussed, with many important questions unaddressed. For example, the single-layer PtCl$_3$ does not have a corresponding bulk layered crystal, so it needs to be fabricated via a bottom-up method. Then, an important question is whether the proposed 2D structure is energetically favored or not. Is there any other competing 2D structural phases? In addition, is there an effective way to control the topological state in 2D PtCl$_3$?

We address these questions in this work. Here, we perform a systematic search of 2D PtCl$_3$ structures, by using the particle swarm optimization (PSO) algorithm combined with the density functional theory (DFT) calculation. The result indicates that the previously proposed CrI$_3$-type structure corresponds to the stable global minimum. Via \emph{ab initio} molecular dynamics simulation, we find that the structure has good thermal stability up to 600 K.  The results suggest that it is possible to realize the desired 2D PtCl$_3$ structure via bottom-up growth method. We also suggest SnS$_2$ as a possible substrate, which has excellent lattice match and preserves the 2D WHS state in PtCl$_3$. We show that uniaxial strains can be used to drive transitions from the WHS state to the QAH state. A moderate strain of 1\% along the armchair direction can open a QAH gap up to 34.9 meV with a unit Chern number. Furthermore, we extend the study to bilayer PtCl$_3$ and find that its magnetic configuration and the electronic structure sensitively depend on the stacking order. Particularly, the $AA'$ stacked bilayer PtCl$_3$ realizes a 2D antiferromagnetic (AFM) mirror Chern insulator, which features a pair of spin-polarized topological edge bands. Our work offers guidance for the experimental studies of 2D PtCl$_3$, reveals interesting interplay among structure, magnetism and band topology in this system, and facilitates the exploration of a variety of novel 2D magnetic topological states.

\section{STRUCTURAL SEARCH}
\begin{figure}[ht]
  \centering
  \includegraphics[width=8cm]{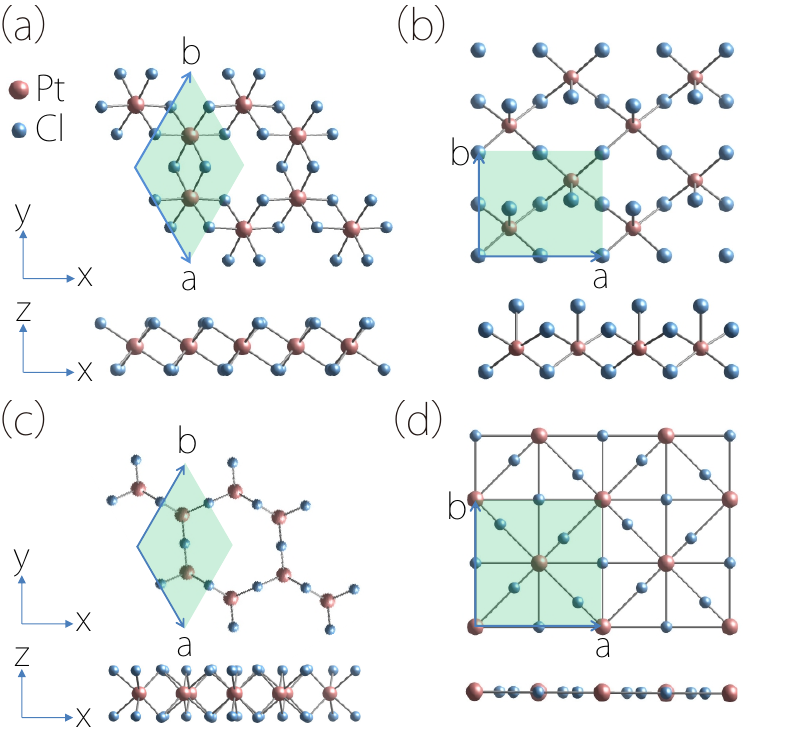}\\
  \caption{(a-d) Top and side views of 2D PtCl$_3$ structures obtained from our structural search. The green shaded region in each figure indicates the primitive unit cell.}\label{fig1}
\end{figure}

We perform a systematic search of 2D PtCl$_3$ structures by using the PSO algorithm as implemented in the CALYPSO package~\cite{wang2010crystal,wang2012calypso,wang2012effective}. The calculation details are presented in Appendix A. The obtained four low-energy structures are plotted in Fig.~\ref{fig1}, labeled as Structure-I to IV. Their key structural information are listed in Table~\ref{tab:lattice}, and the detailed structural data are given in the Supplemental Material.  Our result shows that Structure-I, i.e., the CrI$_3$-type structure proposed in Ref.~\cite{you2019two}, corresponds to the global minimum of 2D PtCl$_3$ structures.

\begin{table}[tb]
	\caption{Information about the four candidate 2D PtCl$_3$ structures.
lattice constants are in unit of \AA. ``Angle'' refers to the angle between $a$ and $b$ in Fig.~1. The energy per unit cell (in unit of eV) is given in reference to Structure-I. }
\label{tab:lattice}
	\begin{tabular}{p{30 pt}p{80pt} p{30pt}p{55pt} cp{30pt}}
		\hline \hline
		      & Lattice constants & Angle & Space group &Energy \\
		\hline
		I     &a=b=6.428    &120$^\circ$& $P$-$31m$          & 0 \\
		II  & a=5.368,b=6.450    &90$^\circ$&   $Pma2$        & 1.527 \\
		III  & a=b=6.644   &121.8$^\circ$&   $Cmmm$      &2.627  \\
         IV  & a=b=7.854  &90$^\circ$&$P4/mbm$           & 5.659 \\
		\hline \hline
	\end{tabular}
\end{table}

For each candidate structure, we investigate its dynamical stability by calculating the phonon spectrum. The result shows that only Structure-I is dynamically stable. The phonon spectra for all other structures in Fig.~\ref{fig1}(b)-(d) contain soft modes (see Supplemental Material)~\cite{SM}. This indicates that Structure-II to IV for PtCl$_3$ cannot exist alone (in a free-standing form), although it does not exclude the possibility that these structures might be stabilized on certain substrates. Due to their instability, we shall focus on Structure-I and not discuss Structure-II to IV further in this work.

\begin{figure}[ht]
  \centering
  \includegraphics[width=8cm]{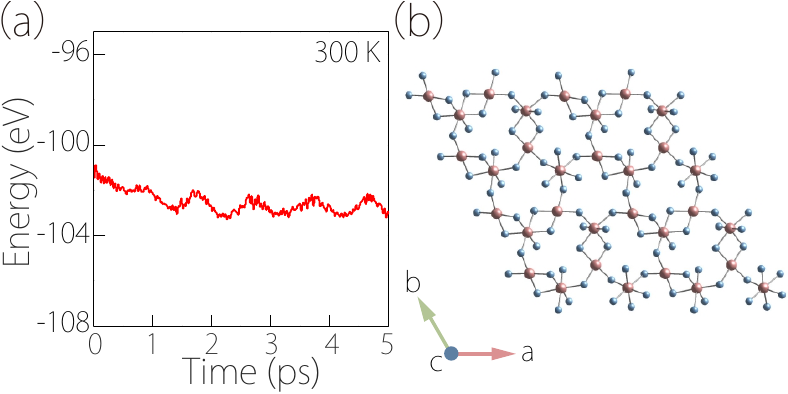}\\
  \caption{(a) Variation of total energy during the AIMD simulation. (b) Snapshot of single-layer PtCl$_3$ lattice at the end of the simulation period.}\label{fig:MD}
\end{figure}

We evaluate the thermal stability of Structure-I by the \emph{ab initio} molecular dynamics simulation (AIMD). The simulated time duration is 5 ps, with a time step of 1 fs. Figure~\ref{fig:MD} shows the variation of total energy versus time and the snapshot of the lattice structure at the end of the simulation duration at 300 K, which indicates that the lattice structure can be well maintained at this temperature. The result for 600 K is similar and is presented in the Supplemental Material.
Our study presented above shows that the single-layer PtCl$_3$ with the CrI$_3$-type structure enjoys a good stability. Other 2D PtCl$_3$ structures are found to have higher energies and are dynamically unstable. Therefore, it is promising to synthesize the desired single-layer PtCl$_3$ structure with WHS state in experiment via certain bottom-up approach, such as molecular beam epitaxy or chemical/physical vapor deposition.

\begin{figure}[ht]
  \centering
  \includegraphics[width=8cm]{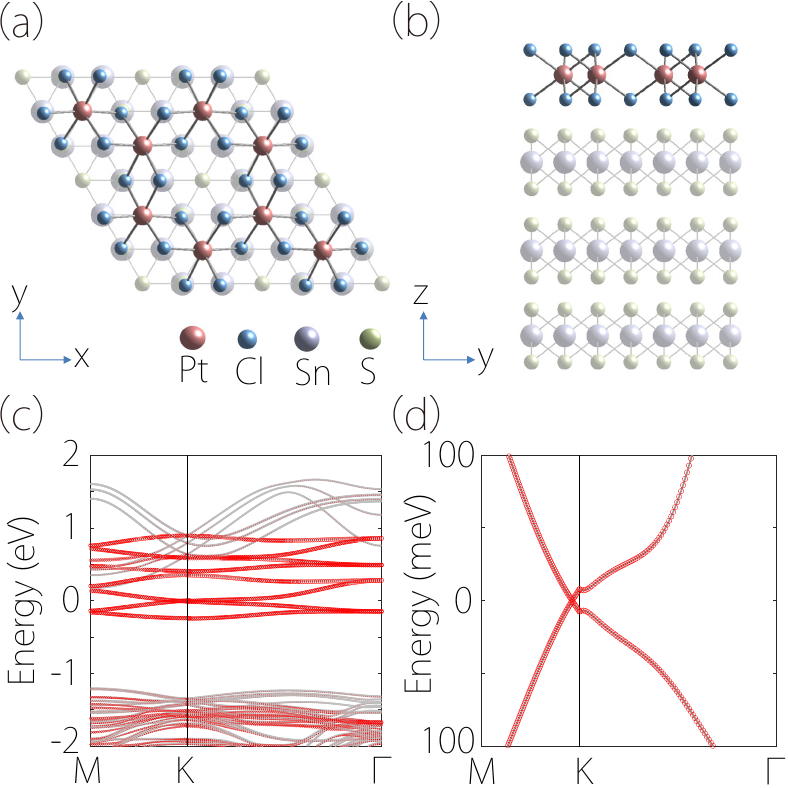}\\
  \caption{(a) Top and (b) side view of 2D PtCl$_3$ on the SnS$_2$ substrate. (c) Band structure of the system.
  The red dot indicates the weight projected onto the PtCl$_3$ layer. (d) Enlarged view of the band structure around the Fermi energy. One observes that the Weyl point of single-layer PtCl$_3$ is preserved. }\label{fig:SnSPtCl$_3$}
\end{figure}

\section{CANDIDATE SUBSTRATE}

For the growth of a 2D material, a suitable substrate is often needed. Here, specific to single-layer PtCl$_3$ (all following discussions are for the CrI$_3$-type structure, i.e., Structure-I, so we will not mention the structure explicitly any more), in order to facilitate experimental study, we must require the substrate to be an insulator which can preserve the WHS character of PtCl$_3$ on top of it.

We have tested several possible substrate materials. For example, MoS$_2$ has a good lattice matching with single-layer PtCl$_3$. The 2 $\times$ 2 supercell of MoS$_2$ has only 1\% mismatch with the primitive cell of PtCl$_3$. However, our calculation shows that a gap opening of 15.8 meV occurs at Weyl points in this case, such that the WHS state in PtCl$_3$ can no longer be preserved. The similar issue was found for the hexagonal boron nitride (hBN) substrate.
For some other substrates, such as GaSe, WSe$_2$, and Si (111), the resulting band structures have other extraneous bands crossing the Fermi level, which may interfere with the WHS physics. More details of our tested substrates are presented in Supplemental Material.

From our test, we find that SnS$_2$ can serve as a suitable substrate. SnS$_2$ is a layered semiconductor, which has been synthesized and received a lot of interest in recent research~\cite{burton2013,ou2019}. The $\sqrt{3}\times\sqrt{3}$ supercell of SnS$_2$ has a negligible mismatch of 0.4\% with the primitive cell of PtCl$_3$.
Figure~\ref{fig:SnSPtCl$_3$}(a,b) illustrates the optimized structure of single-layer PtCl$_3$ on top of the SnS$_2$ substrate. In the model, we take the substrate thickness to be three layers of SnS$_2$.
The calculated band structure of the system is plotted in Fig.~\ref{fig:SnSPtCl$_3$}(c)-(d). One can see that the low-energy states near the Fermi level are from PtCl$_3$, which sit inside the band gap of the SnS$_2$ substrate. From Fig.~\ref{fig:SnSPtCl$_3$}(d), it is clear that the Weyl point is preserved in the presence of the substrate. Our result suggests that SnS$_2$ could be a promising substrate for the growth of single-layer PtCl$_3$.

\begin{figure*}[ht]
  \centering
  \includegraphics[width=14cm]{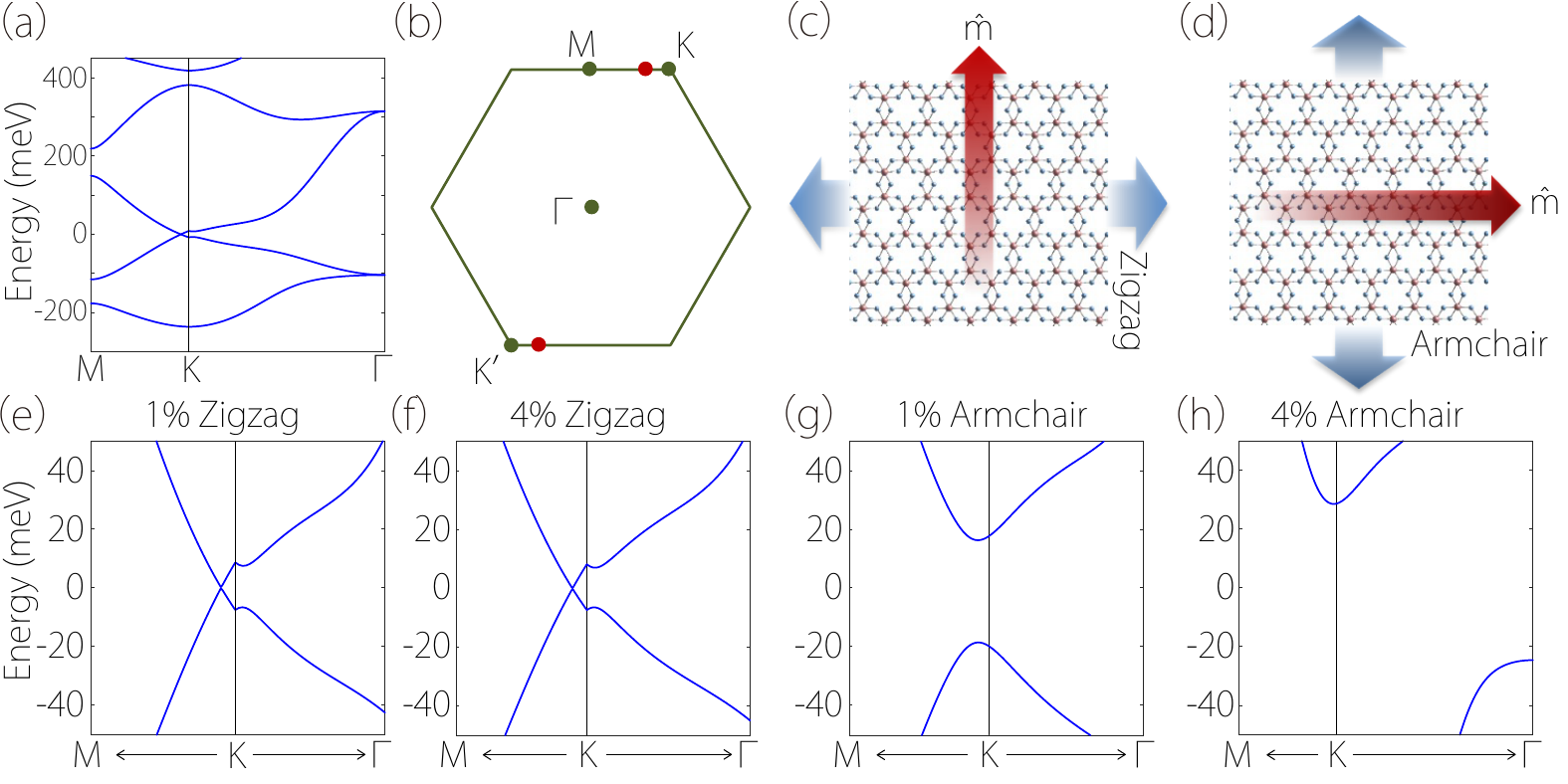}\\
  \caption{(a) Brillouin zone and (b) band structure for the unstrained single-layer PtCl$_3$. The red dots in (a) illustrate the location of the two fully spin-polarized Weyl points. (c) The magnetic easy axis is still along the armchair direction when unaxial strain is applied in the zigzag direction. (d) The easy axis rotates to the zigzag direction for uniaxial strain ($>0.5\%$) applied along the armchair direction. In (c) and (d), the blue arrows indicate the strain direction. (e-f) Band structures for the strained single-layer PtCl$_3$. (e-f) For uniaxial strain along the zigzag direction, the WHS state is preserved. (g-h) Transition to QAH insulator state occurs for uniaxial strain applied along the armchair direction.}\label{fig:strain}
\end{figure*}

\section{STRAIN-induced QAH PHASE}

Two-dimensional materials typically have much better flexibility than 3D materials. Lattice strain can be readily applied for 2D materials in experiment. For example, a well developed technique is to transfer the 2D material after growth onto a stretchable substrate, and then
strain is applied in a controllable way to the material by stretching the substrate~\cite{conley2013bandgap}. Using this technique, strains \textgreater 10\% have been demonstrated in experiment~\cite{kim2009large}.

Strain could produce interesting effects in single-layer PtCl$_3$. In the ground state, single-layer PtCl$_3$ is a ferromagnet with its magnetization direction $\hat{\bm m}$ along an easy-axis in the $x$-$y$ plane. In Ref.~\cite{you2019two}, it was shown that this easy-axis is normal to a vertical mirror plane (or a mirror line in 2D) of the lattice, such that the mirror symmetry is preserved by the magnetism. In the coordinate system in Fig.~\ref{fig1}(a), $\hat{\bm m}$ could be, for example, along the $y$ direction, which is the armchair direction with respect to the Pt sublattice, then the mirror symmetry $M_y$ is preserved for the system.
The WHS state is protected by this mirror symmetry. The two Weyl points are located on the mirror-invariant $K$-$K'$ path of the Brillouin zone (BZ). The case for $\hat{\bm m}=\hat{y}$ is illustrated in Fig.~\ref{fig:strain}(a,b). The Weyl points are protected, because the two crossing bands on this path have opposite $M_y$ eigenvalues.

Now, we study the effects of unaxial strains on single-layer PtCl$_3$. We shall only consider tensile strains, as they are more easily applied to 2D materials in experiment. Let's first consider applied strain along the $x$ direction in Fig.~\ref{fig1}(a), which is the zigzag direction with respect to the Pt sublattice. We find that the direction of magnetization $\hat{\bm m}$ still prefers the $y$ direction, as illustrated in Fig.~\ref{fig:strain}(c). It follows that the $M_y$ symmetry is preserved, and hence the Weyl points (and the WHS state) are maintained. This is confirmed by the band structure results in Fig.~\ref{fig:strain}(e) and (f).

On the other hand, uniaxial strains along the $y$ direction, i.e., the armchair direction, will give a different result. We find that under a small strain $>0.5\%$, the magnetic easy axis will change to zigzag direction, as illustrated in Fig.~\ref{fig:strain}(d). Consequently, the original $M_y$ symmetry is broken. Then, the Weyl points lose the protection and will generally be gapped out. Our calculation confirms this point. As shown in Fig.~\ref{fig:strain}(g), under $1.0\%$ strain, the Weyl point is destroyed, and a gap $\sim 34.9$ meV is opened in the spectrum. This represents a strain-induced semimetal to semiconductor phase transition. Interestingly, different from conventional scenarios where the transition is due to the modification of band dispersion (by strain) which pushes the conduction and valence bands apart;  here, the transition is enabled by the change in magnetic anisotropy, which breaks the protecting symmetry of the band crossing Fermi points.

\begin{figure}[ht]
  \centering
  \includegraphics[width=8cm]{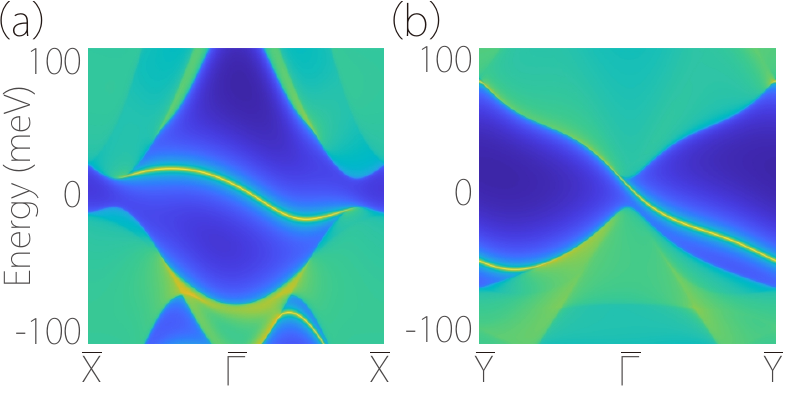}\\
  \caption{The edge spectrum corresponding to the case in
  Fig.~\ref{fig:strain}(g) for the edge along (a) zigzag and (b) armchair direction. The gapless chiral edge band for the QAH state can be observed. }\label{Fig:EdgeStrainXY}
\end{figure}

More interestingly, the resulting semiconductor state after transition turns out to be a QAH state. For the band structures in Fig.~\ref{fig:strain}(g) and (h), our calculation shows that the valence bands are characterized by a unit Chern number $|\mathcal{C}|=1$ (the sign depends on whether $\hat{\bm m}$ is along $+\hat{x}$ or $-\hat{x}$). The Chern number is related to the integral of valence-band Berry curvature field over the BZ. Here, the Berry curvature is concentrated in the regions around the two gapped Weyl points. Each point contributes about a half of the Chern number.
The unit Chern number requires that on the edge of the 2D system, there must be one gapless chiral edge band crossing the bulk gap, which is confirmed by the calculated edge spectra in Fig.~\ref{Fig:EdgeStrainXY}. These results demonstrate that applying strain is an effective way to tune the topological states in single-layer PtCl$_3$.

\begin{figure*}[ht]
  \centering
  \includegraphics[width=16cm]{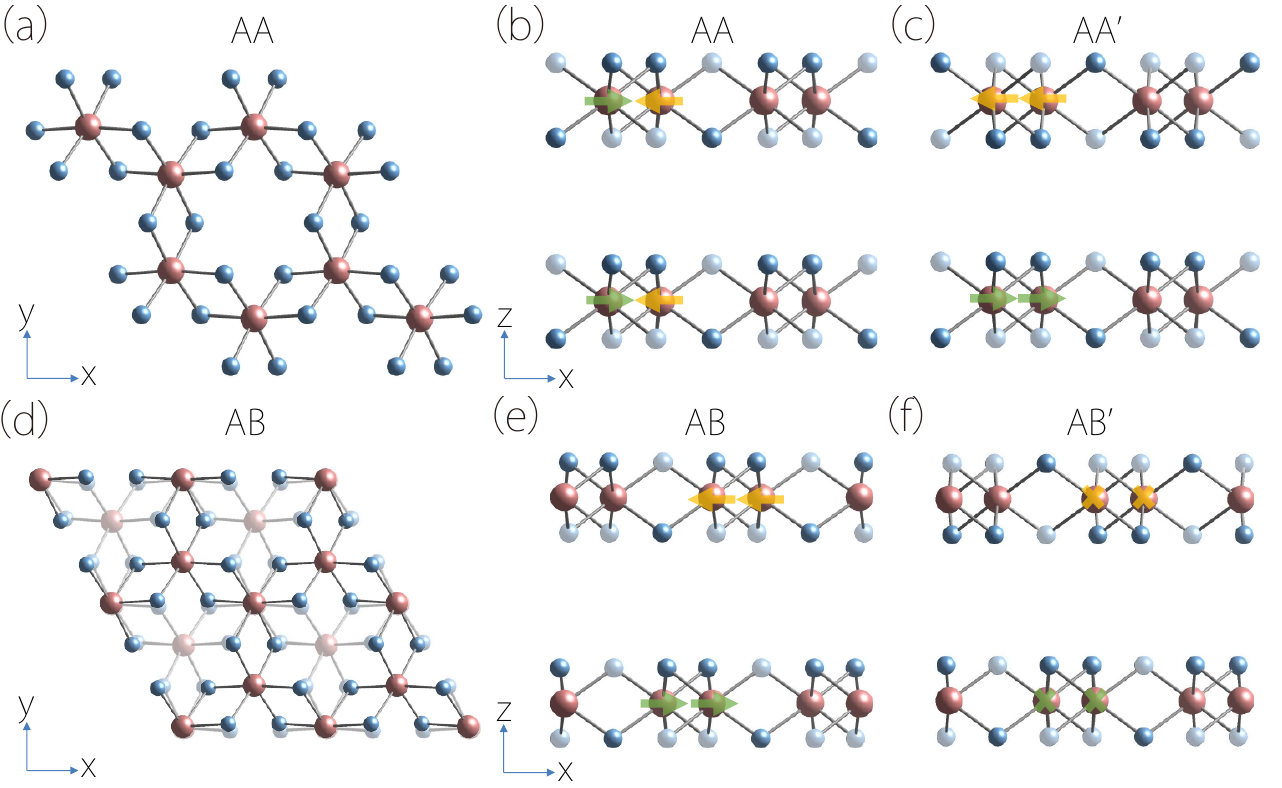}\\
  \caption{Crystal structures for different stacking configurations of bilayer PtCl$_3$. (a) Top and (b) side view of $AA$ stacking. (c) $AA'$ stacking is obtained from $AA$ stacking by imposing the horizontal mirror reflection $M_z$ on the top layer. In the side view, some Cl atoms are made into transparent color style, in order to indicate that they are at different $x$ coordinates as those Cl atoms without transparency. Comparing (b) and (c), one can see that $AA'$ stacking preserves an $M_z$ symmetry, but $AA$ stacking does not.
  (d) Top and (e) side view of $AB$ stacking. (f) Like $AA'$ stacking, $AB'$ stacking is obtained from $AB$ stacking by flipping the top layer upside down.
  The arrows in the figures indicate the local spin direction. The green arrows are opposite to the yellow arrows. In (f), the local spins are in the $x$ direction.}\label{fig:bilayer}
\end{figure*}

\section{magnetism in bilayers}

In the past ten years, there has been rapid development in the experimental techniques to construct van der Waals layered structures via stacking 2D materials~\cite{novoselov2016}. Nowadays, the stacking configuration can be well controlled, and even fine structural tunings, such as twisting two layers with a tiny angle, have been demonstrated in experiment~\cite{cao2018correlated}. It has been found that different stacking orders may have strong impacts on the material properties.
In the following, we extend our study to bilayer PtCl$_3$ and show that the stacking configuration can have important effects on the magnetism of the system.

We consider four representative stacking configurations, as illustrated in Fig.~\ref{fig:bilayer}. The simplest one is the $AA$ stacking, for which the top layer is just an identical copy of the bottom layer, shifted up along the $z$ direction. Since in each single layer the Pt atoms form a honeycomb lattice, there is a natural $AB$ stacking obtained by a lateral shift of the top layer in $AA$ stacking, such that one triangular Pt sublattice in the top layer is above the center of the Pt hexagon in the bottom layer (and vice versa), just like the $AB$ stacking in bilayer graphene. In addition, one notes that single-layer PtCl$_3$ does not have a horizontal mirror plane (it has roto-reflection $S_6$), so we should also consider the case when one of the layer, e.g., the top layer, is flipped upside down. Starting from $AA$ and $AB$ stackings, this operation generates another two
configurations, which we denote as $AA'$ and $AB'$, respectively. For each stacking, we have optimized its lattice structure. The obtained interlayer distances are listed in Table \ref{tab:magnet}.

We have investigated the magnetic ground state for each stacking configuration. The results are presented in Table \ref{tab:magnet}. One observes that for each stacking considered, the bilayer PtCl$_3$ would favor some kind of AFM ordering. However, the details depend on the stacking. As shown in Fig.~\ref{fig:bilayer}(c) and (e), for $AA'$ and $AB$ stacking, each PtCl$_3$ layer exhibits a ferromagnetic ordering with local spins along the $y$ direction, i.e., perpendicular to the mirror line, which inherits the feature of the single-layer case. Meanwhile, the interlayer magnetic coupling is of AFM type. The result for $AB'$ stacking is similar: the coupling is of ferromagnetic type within each layer and of AFM type between the two layers, but the easy-axis is along $x$, i.e., along the mirror line. Finally, for $AA$ stacking, the intralayer coupling is of AFM type, namely, the two triangular lattices have opposite local spins, along the $y$ direction, and the interlayer coupling between each vertical pair of Pt sites is ferromagnetic.

In the following section, we shall see that the different magnetic ordering corresponding to different stacking will in turn strongly affect the electronic band structure of bilayer PtCl$_3$.

\begin{table}[ht]
	\caption{Information for bilayer PtCl$_3$ with different stacking. FM stands for the ferromagnetic ordering. AFM$_1$ refers to the configuration with intralayer AFM ordering. AFM$_2$ refers to the configuration with intralayer FM ordering and interlayer AFM ordering. The energy values (in unit of meV per unit cell) are with reference to the ground state of each stacking.
MSG stands for the magnetic space group. $d$ (in unit of \AA) is the distances between the layers.
 }
\setlength{\tabcolsep}{1.25mm}
\label{tab:magnet}
	\begin{tabular}{lcccccc}
		\hline \hline
		                     & FM & AFM$_{1}$ & AFM$_{2}$ & Easy axis & MSG & $d$\\
		\hline
		$AA$  &438    &0      &440        &y    & C2/m$^{\prime}$(12.61) & 4.44 \\
		$AA^\prime$  &2    &92          &0      &y    &Amm2(38.187) & 4.38 \\
$AB$     &662    &720    &0       &y    & P-1$^{\prime}$(2.6) & 4.94 \\
$AB^{\prime}$  &601    &223      &0          &x    &C2$^{\prime}$(5.15) & 4.95 \\
		\hline \hline
	\end{tabular}
\end{table}

\section{AFM Mirror Chern insulator}

\begin{figure}[tb]
  \centering
  \includegraphics[width=\linewidth]{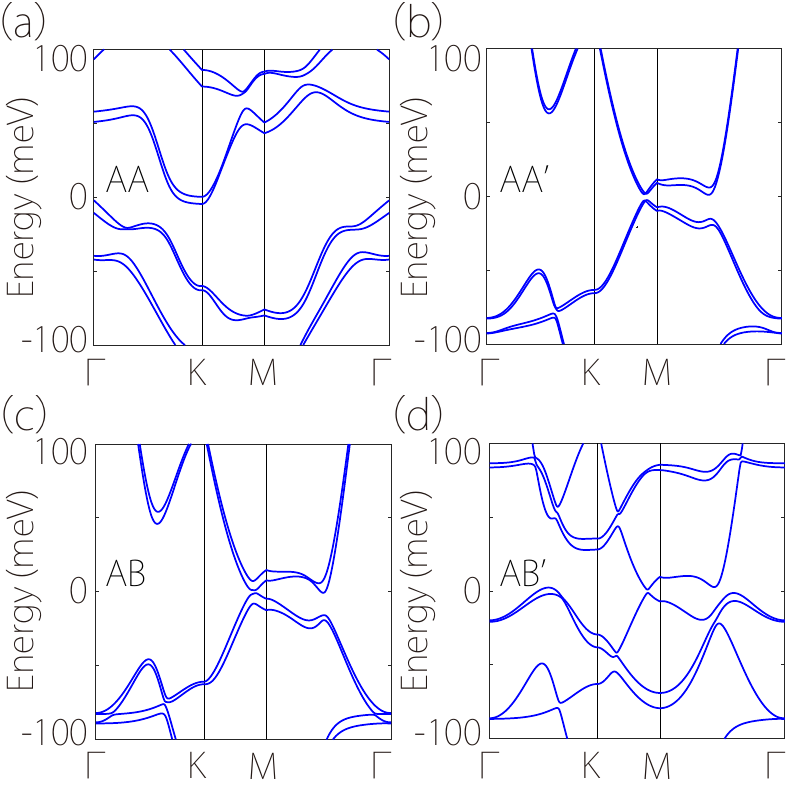}\\
  \caption{Band structure of bilayer PtCl$_3$ with (a) $AA$, (b) $AA'$, (c) $AB$, and (d) $AB'$ stacking.  Magnetism and SOC are fully included. }\label{fig:bilayerband}
\end{figure}

We have calculated the electronic band structures for the four stacking configurations. The results are plotted in Fig.~\ref{fig:bilayerband}. One can see that $AA$ and $AB'$ stacking corresponds to (semi)metallic states [Fig.~\ref{fig:bilayerband}(a) and (d)], whereas the other two are AFM semiconductors [Fig.~\ref{fig:bilayerband}(b) and (c)]. The band gaps for $AA'$ and $AB$ stacking are $\sim 4.5$ meV and $\sim 2.1$ meV, respectively. One notes that the band structures for $AA'$ and $AB$ are similar, owing to their similar magnetic ordering. The result reflects a strong coupling among layer stacking, magnetism, and electronic excitation in bilayer PtCl$_3$ system.

We have seen that single-layer PtCl$_3$ hosts interesting magnetic topological states: It is a WHS in the ground state and can be converted into a QAH insulator by applied magnetic field or strain. It is natural to ask whether the bilayer structures have any nontrivial topology. From Fig.~\ref{fig:bilayerband}, one observes that the original Weyl points in single-layer PtCl$_3$ has been destroyed, due to the interlayer interaction. For $AA$, $AB$, and $AB'$ stacking, we find that their band structures are topologically trivial. On the other hand, $AA'$ stacking features an interesting topological state --- the AFM mirror Chern insulator. We shall focus on
$AA'$ stacked bilayer PtCl$_3$ in the following discussion.

\begin{figure}[ht]
  \centering
  \includegraphics[width=8cm]{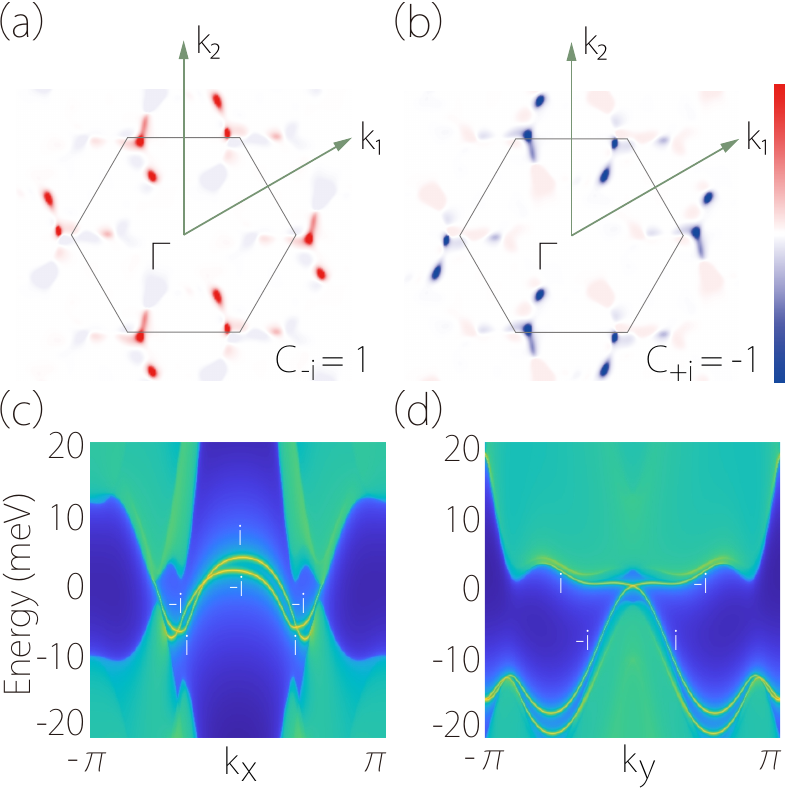}\\
  \caption{(a,b) Berry curvature distribution for the two mirror subspaces.   (e,f) Edge spectra for (e) zigzag edge and (f) armchair edge, which exhibit a pair of gapless edge bands corresponding to the AFM mirror Chern insulator state. }\label{Fig:AApy}
\end{figure}

First, we note that the $AA'$ stacked bilayer PtCl$_3$ possesses the horizontal mirror plane symmetry $M_z$. Under this mirror reflection, the top layer is mapped to the bottom layer, and vice versa. It is important to note that the AFM ordering in $AA'$ preserves $M_z$: The in-plane local spins are reversed under $M_z$, since they are pseudo-vectors. Because all the $k$ points in the 2D BZ are invariant under $M_z$, the band eigenstates can be divided into two subspaces according to their $M_z$ eigenvalues. The mirror Chern number for the 2D system can be defined as~\cite{teo2008surface}
\begin{equation}
  \mathcal{C}_{M_z}=\frac{1}{2}\left(\mathcal{C}^{M_z}_+ - \mathcal{C}^{M_z}_-\right),
\end{equation}
where $\mathcal{C}^{M_z}_\pm$ is the Chern number for valence bands with $\pm$ eigenvalue of $M_z$. We have evaluated these Chern numbers for the first-principles band structure in Fig.~\ref{fig:bilayerband}(b) and obtained $\mathcal{C}^{M_z}_+=-\mathcal{C}^{M_z}_-=-1$. (The Berry curvature distribution for the two subspaces are plotted in Fig.~\ref{Fig:AApy}(a,b)). The result indicates that the system can be regarded as two copies of Chern insulators with opposite (unit) Chern numbers. The two copies are decoupled by the $M_z$ symmetry, so they will not annihilate with each other. In previous works, mirror Chern insulators were discussed mostly as subsystems of 3D materials and for non-magnetic cases~\cite{hsieh2012topological,tanaka2012experimental}. In comparison, here, we have a magnetic mirror Chern insulator realized in a concrete 2D material with AFM ordering.

The nontrivial mirror Chern number leads to topological edge modes at the boundary of $AA'$ stacked bilayer PtCl$_3$. Since $\mathcal{C}^{M_z}_+=-\mathcal{C}^{M_z}_-=-1$, we should have one gapless edge band for each mirror subspace, and the two edge bands should have opposite chirality. This is confirmed by our calculation result, as shown in Fig.~\ref{Fig:AApy}(c,d) (in the figure, we have labeled the $M_z$ eigenvalues for the edge bands).

\begin{figure}[ht]
  \centering
  \includegraphics[width=8cm]{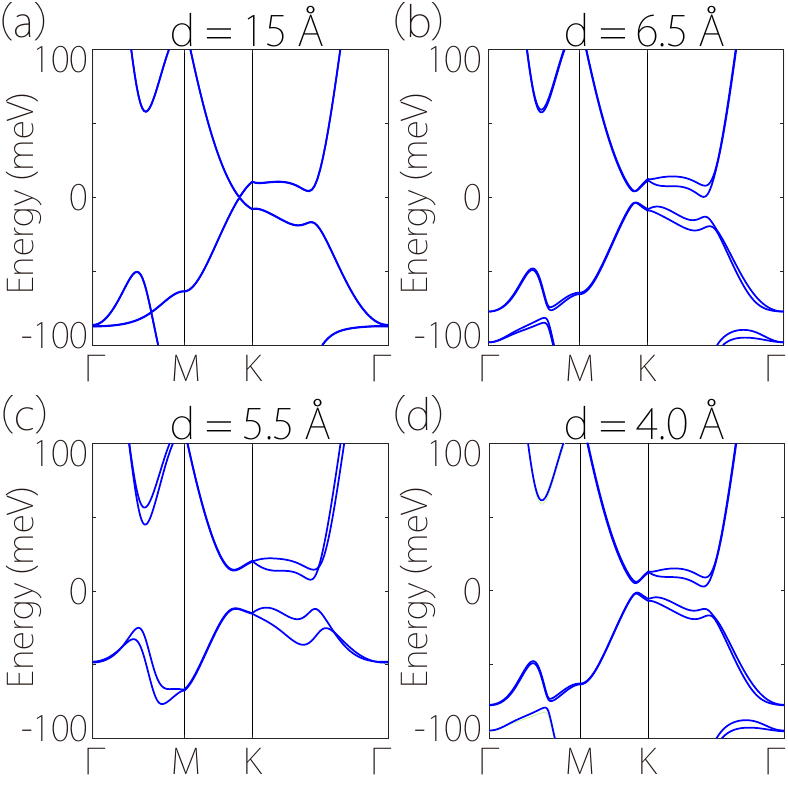}\\
  \caption{(a-d) Band structures of $AA'$ stacked bilayer PtCl$_3$ with different interlayer distance $d$. For the gapped cases in (b-d), the system is in the AFM mirror Chern insulator state.}\label{fig9}
\end{figure}

To better understand the formation of the AFM mirror Chern insulator state, we investigate the evolution of band structure when two PtCl$_3$ single layers are brought together to form a bilayer. In Fig.~\ref{fig9}, we plot the band structure (for $AA'$ stacking) at four different interlayer spacings. One observes that when the two layers are largely separated [Fig.~\ref{fig9}(a)] and the interlayer interaction is negligible, the total band structure just consists of two copies of that for a single-layer. The nodal point in Fig.~\ref{fig9}(a) is not a single Weyl point but two Weyl points overlying onto each other. Such a degeneracy is not stable and will be gapped out by the interaction between the two layers. The gap becomes obvious with the enhanced interlayer interaction at decreased interlayer separation [Fig.~\ref{fig9}(b-d)]. For example, at $d=5.5$ \AA, the band gap can be $\sim 18.9$ meV. Note that single-layer PtCl$_3$ features band inversion, so the gap opening will drive it into a QAH state. If we treat the interlayer coupling as a perturbation, then each layer can be viewed as a QAH insulator, but the two have opposite Chern numbers due to their opposite magnetization direction.
As the net Chern number cancels out, typically, the composite system would become trivial. Fortunately, the mirror symmetry $M_z$ comes to the rescue. It ensures a nontrivial mirror Chern number and protects the crossing points between the chiral edge bands.

\section{Discussion and Conclusion}

So far, single-layer PtCl$_3$ is the only material candidate for the 2D WHS state. We hope our current study can facilitate its experimental synthesis, e.g., through molecular beam epitaxy or chemical vapor deposition growth. The revealed strain-driven topological phase transition not only offers a new route to QAH insulators, it may also be utilized for designing new topological electronic devices, such as strain sensors.

We have revealed rich physics in bilayer PtCl$_3$, originated from the coupling among stacking, magnetism, and electronic band structure. Particularly, the $AA'$ stacked bilayer realizes a previously unknown AFM mirror Chern insulator state.
It could be even more interesting if one can achieve twisted bilayer PtCl$_3$. Generally, a small twist angle will lead to a moire superlattice with large period~\cite{li2010observation,bistritzer2011moire,ribeiro2018twistable}. Different regions in the cell will correspond to different local stacking types.
Since we have shown that in bilayer PtCl$_3$, the magnetic ordering and electronic band structure have sensitive dependence on the stacking, one can expect that in the twisted bilayer, there may emerge periodic magnetic textures, topological domains, and etc., which are interesting topics to explore in future research.

In conclusion, by first-principles calculations and theoretical analysis, we have clarified several important questions about single-layer and bilayer  PtCl$_3$. For single layer, via extensive structural search using the PSO algorithm, we show that the proposed PtCl$_3$ structure (which hosts the WHS state) is energetically favored and enjoys good thermal stability. We find that SnS$_2$ could be a suitable substrate for 2D PtCl$_3$, due to its negligible lattice mismatch and preserving the WHS state in PtCl$_3$.
We also show that uniaxial strains along the armchair direction can drive a topological phase transition into the QAH state, whereas strains along the zigzag direction preserve the WHS state. As for bilayer PtCl$_3$, we show that the magnetic ordering and electronic structure strongly depends on the stacking configuration. Notably, the $AA'$ stacking gives rise to a novel 2D AFM mirror Chern insulator state. Our work provides guidance for the experimental realization of 2D PtCl$_3$ and reveals its potential as a platform to study various magnetic topological states.

\begin{acknowledgements}
We thank D. L. Deng for helpful discussions. This work is supported by the NSFC (Grants No. 12174018, No. 12074024, No. 11774018), and the Singapore Ministry of Education AcRF Tier 2 (T2EP50220-0026). We acknowledge computational support from the Texas Advanced Computing Center.
\end{acknowledgements}

\ \
\par
\ \
\begin{appendix}
\renewcommand{\theequation}{A\arabic{equation}}
\setcounter{equation}{0}
\renewcommand{\thefigure}{A\arabic{figure}}
\setcounter{figure}{0}
\renewcommand{\thetable}{A\arabic{table}}
\setcounter{table}{0}

\section{First-principles method}

The structural search was performed by using the CALYPSO package. The population size was set to 30. Unit cells containing $1$ to  $4$ times of the formula unit were considered. 60\% of the searched structures in each generation were evolved into the next generation by PSO, and the other 40\% were randomly generated. The buckled structures were considered by setting the control parameter of buffering thickness to 0.6 Å. Subsequent structural relaxation was performed on the basis of DFT, using the generalized gradient approximation (GGA) in the form proposed by Perdew, Burke and Ernzerhof (PBE)~\cite{PRL3865} as implemented in the Vienna ab initio Simulation Package (VASP)~\cite{PRB11169,kresse1996efficient}. The energy cutoff of the plane-wave was set to 500 eV. The Grimme dispersion correction~\cite{grimme2006semiempirical} was used to account for the van der Waals interaction. The structures were fully relaxed until the maximum force on each atom was less than 0.01 eV/\AA. The energy convergence criterion in the self-consistent calculations was set to 10$^{-5}$ eV. A Gamma-centered Monkhort-Pack $k$-point mesh with a resolution of $2 \pi \times 0.03 $ Å was used for geometry optimization and self-consistent calculations. A vacuum slab of 15 Å in the $z$ direction was adopted to avoid artificial interactions between neighboring layers. To account for possible correlation effects for Pt $d$ orbitals, the GGA+$U$ method~\cite{anisimov1991band} with $U=1 eV$ was used for calculating the band structures. The phonon dispersion was computed by using the Phonopy code~\cite{togo2008first} within the density functional perturbation theory~\cite{gonze1997dynamical}. In phonon calculations, a finer $k$-point grid of $2\pi \times 0.02 $ Å was employed. The edge spectrum was calculated by using the Wannier functions and the iterative Green's function method~\cite{marzari1997maximally,wu2018wanniertools,sancho1984quick}.

\end{appendix}

\bibliographystyle{apsrev4-2}

\bibliography{PtCl}

\end{document}